\documentclass[%
 aip,
 jmp,%
 amsmath,amssymb,
 reprint,%
]{revtex4-1}

\usepackage{graphicx}
\usepackage{dcolumn}
\usepackage{bm}

\begin{document}

\preprint{AIP/123-QED}

\title{Non-parametric reconstruction of dark energy and cosmic expansion from the Pantheon compilation of type Ia supernovae}

\author{Hai-Nan Lin}
 \email{linhn@ihep.ac.cn}
\affiliation{Department of Physics, Chongqing University, Chongqing 401331, China}
\author{Xin Li}
\email{lixin1981@cqu.edu.cn}
 \affiliation{Department of Physics, Chongqing University, Chongqing 401331, China}
\author{Li Tang}%
 \email{tang@cqu.edu.cn}
\affiliation{Department of Physics, Chongqing University, Chongqing 401331, China}
 %


\begin{abstract}
The equation of state (EoS) of dark energy plays an important role in the evolution of the universe and arouses great interests in recent years. With the progress on observational technique, precise constraint on the EoS of dark energy becomes possible. In this paper, we reconstruct the EoS of dark energy and cosmic expansion using Gaussian processes (GP) from the most up-to-date Pantheon compilation of type Ia supernovae (SNe Ia), which consists of 1048 finely calibrated SNe Ia. The reconstructed EoS of dark energy has large uncertainty due to its dependence on the second order derivative of the construction. Adding the direct measurements of Hubble parameters $H(z)$ as an additional constraint on the first order derivative can partially reduce the uncertainty, but is still not precise enough to distinguish between evolving and constant dark energy. Besides, the results heavily rely on the prior of Hubble constant $H_0$. The $H_0$ value inferred from SNe+$H(z)$ without prior is $H_0=70.5\pm 0.5~{\textrm{km}~\textrm{s}^{-1}~\textrm{Mpc}^{-1}}$. Moreover, the matter density $\Omega_M$ has an unnegligible effect on the reconstruction of dark energy. Therefore, more accurate determinations on $H_0$ and $\Omega_M$ are needed to tightly constrain the EoS of dark energy.

\end{abstract}

\maketitle

\section{Introduction}
The late time cosmic acceleration is one of the most important discoveries in modern cosmology and it revives Einstein's cosmological constant hypothesis. Since the first discovery of cosmic acceleration from type Ia supernovae (SNe Ia) in the late 1990s \cite{Perlmutter:1999,Riess:1998}, it has now been confirmed by various other independent observations such as the large scale structure \cite{Tegmark:2004}, the growth function \cite{Nesseris:2008} and the cosmic microwave background radiations \cite{Ade:2015xua,Aghanim:2018eyx}. This leads to the final foundation of the standard model of cosmology, i.e. the cold dark matter plus a cosmological constant ($\Lambda$CDM) model. Here $\Lambda$ stands for the cosmological constant (or an alternative name dark energy), which provides a negative pressure and is responsible for the acceleration of the universe. According to the $\Lambda$CDM model, the equation of state (EoS) of dark energy is a constant and does not evolve with cosmos, i.e. $w=p/\rho\equiv -1$. Although the $\Lambda$CDM model has achieved great success, it still confronts some problems, among which the most important ones are the coincidence problem and fine-turning problem \cite{Weinberg:1989,Zlatev:1998tr}. The EoS of dark energy plays an essential roles in the evolution of universe. In terms of different EoS of dark energy, several alternative models have been proposed, such as the model with a constant $w$ but does not necessarily equate to $-1$, the evolving dark energy models e.g. the Chevallier-Polarski-Linder parametrization \cite{Chevallier:2001,Linder:2003} and various other parameterizations \cite{Yang:2017alx,Xu:2016grp}. In some models the dark energy is replaced by a scalar field, such as the quintessence field \cite{Caldwell:1988}, phantom field \cite{Caldwell:2002}, tachyon field \cite{Sen:2002}, etc. The effective EoS of these scalar fields is also evolving with cosmos.

The above models depend on the specific parametrization of dark energy or the scalar fields and thus are model dependent. Moreover, most of the parameterizations are lack of physical interpretation hence are just phenomenological. Since we have no prior knowledge on the explicit form of dark energy, reconstructing it in a non-parametric way is of great importance. To this end, some model-independent methods have been proposed, among which the Gaussian processes (GP) is one of the most widely used methods. Unlike the best-fitting method which must have a concrete model to fit the data, the GP method can reconstruct a theoretical curve from the discrete data points without evolving any specific model. Since Ref.\cite{Holsclaw:2010sk} first applied the GP method to investigate the dark energy, it has been widely used and has shown its powerful ability in cosmology \cite{Holsclaw:2010nb,Holsclaw:2011wi,Seikel:2012uu,Zhang:2016tto,Zhang:2018gjb,Yin:2018mvu,Aghamousa:2017uqe,Gonzalez:2016lur,Cai:2015zoa,Cai:2015pia}. The advantage of GP method is that it does not need the concrete form of the model, the only assumption is that the observational data points are drawn from the multivariate Gaussian distribution.

In this paper, we try to reconstruct the EoS of dark energy using the GP method from the latest dataset of SNe Ia, i.e. the Pantheon compilation \cite{Scolnic:2017caz}, which consist of 1048 finely calibrated SNe Ia. The EoS of dark energy has strong influence on the Hubble expansion rate $H(z)$ and the deceleration parameter $q(z)$, which will be obtained simultaneously in the reconstruction. The reconstructed dark energy depends on the second order derivative of GP (see the next section for details), hence has large uncertainty. To improve the significance, we will use the direct measurement of Hubble parameters $H(z)$ as an additional constraint in the GP reconstruction. Since there are more than $3\sigma$ tension between the Hubble constant $H_0$  from the local distance ladders \cite{Riess:2016jrr} and from the global CMB radiation \cite{Aghanim:2018eyx}, we will also investigate the impact of different values of $H_0$ on the reconstruction.

The rest of the paper is organized as follows: In section 2, we introduce the methodology and the relevant data that are used in our analysis. The results together with some discussions are presented in section 3. Finally, a short summary is given in section 4.

\section{Data and methodology}\label{sec:methodology}

The Hubble expansion rate $H(z)$ strongly depends on the contents of the universe and the EoS of dark energy. In a spatially flat Friedmann-Robertson-Walker universe dominated by non-relativistic matter (including baryons and dark matter) and dark energy, the evolution of $H(z)$ is governed by the Friedmann equation \cite{Weinberg:2008}
\begin{equation}\label{eq:hubble}
  H^2(z)=H_0^2\left\{\Omega_M(1+z)^3+\Omega_\Lambda\exp\left[3\int_0^z\frac{1+w(z)}{1+z}dz\right]\right\},
\end{equation}
where $H_0$ is the Hubble constant, $\Omega_M$ and $\Omega_\Lambda$ are respectively the normalized density of non-relativistic matter and dark energy at current epoch, and $w(z)=p(z)/\rho(z)$ is the EoS of dark energy. The normalized comoving distance is related to the Hubble expansion rate by \cite{Hogg:1999ad}
\begin{equation}\label{eq:comoving}
  d_c(z)=\int_0^z\frac{dz}{E(z)},
\end{equation}
where $E(z)\equiv H(z)/H_0$ is the normalized Hubble parameter. From equation (\ref{eq:comoving}) we have
\begin{equation}\label{eq:E(z)}
  E(z)=\frac{1}{d_c'(z)},
\end{equation}
where the prime denotes the derivative with respect to redshift $z$. Combining equations (\ref{eq:hubble}) and (\ref{eq:E(z)}) we can solve for $w(z)$,
\begin{equation}\label{eq:w(z)}
  w(z)=\frac{-2(1+z)d_c''-3d_c'}{3[d_c'-\Omega_M(1+z)^3d_c'^3]}.
\end{equation}

The acceleration of the universe is often represented by the so-called ``deceleration parameter", which is defined by $q(z)=-\ddot{a}a/\dot{a}^2$, where $a=1/(1+z)$ is the scale factor of the universe, and the dot denotes the derivative with respect to cosmic time. Using $H=\dot{a}/a$, the deceleration parameter can be rewritten as
\begin{equation}\label{eq:q(z)}
  q(z)=(1+z)\frac{H'}{H}-1=-(1+z)\frac{d_c''}{d_c'}-1.
\end{equation}
A positive or negative $q$ means a decelerating or accelerating universe, respectively.

If we know $d_c(z)$ as the function of $z$, then $E(z)$ can be obtained from equation (\ref{eq:E(z)}), similarly $w(z)$ and $q(z)$ can be obtained from equations (\ref{eq:w(z)}) and (\ref{eq:q(z)}), respectively. In a spatially flat universe, the normalized comovig distance $d_c(z)$ is related to the luminosity distance $D_L(z)$ by
\begin{equation}
  d_c(z)=\frac{1}{1+z}\frac{H_0}{c}D_L(z).
\end{equation}
The luminosity distance can be measured from the brightness of SNe Ia. SNe Ia are often assumed to have an approximately constant absolute magnitude after the color and stretch corrections so are widely regarded as the standard candles. The distance modulus of SNe Ia can be derived from the observation of light curves through the empirical relation \citep{Tripp:1998,Guy:2005,Guy:2007}
\begin{equation}\label{eq:mu_sn}
  \mu_{\textrm{sn}}=m_B^*+\alpha X_1-\beta \mathcal{C}-M_B,
\end{equation}
where $m_B^*$ is the B-band apparent magnitude, $X_1$ and $\mathcal{C}$ are the stretch and color parameters respectively, and $M_B$ is the absolute magnitude. $\alpha$ and $\beta$ are two nuisance parameters. The luminosity distance of SNe Ia can be calculated from the distance modulus through the following relation
\begin{equation}
  \mu=5\log_{10}\frac{D_L}{\textrm{Mpc}}+25.
\end{equation}

Several SNe Ia samples have been released, among which the most up-to-date one is the Pantheon compilation \cite{Scolnic:2017caz}. The Pantheon sample is at present the largest sample which consists of different supernovae surveys, including SDSS, SNLS, various low-z samples and some high-z samples from HST. The total number of SNe in the Pantheon dataset is 1048, which is about twice of the Union2.1 sample \cite{Suzuki:2012}, and is about $40\%$ more than the JLA sample \cite{Betoule:2014frx}. Moreover, the furthest SNe reaches to $z\sim 2.3$ and the systematic uncertainty is further reduced compared to the previous samples. Usually, the nuisance parameters $\alpha$ and $\beta$ are optimized simultaneously with the cosmological parameters or are marginalized over. However, this method is model dependent thus the distance calibrated in one cosmological model couldn't be directly used to constrain the other models. The Pantheon sample applies a new method called BEAMS with Bias Corrections (BBC) to calibrated the SNe. According to the BBC method, the nuisance parameters $\alpha$ and $\beta$ are determined by fitting to an randomly chosen reference cosmology. There is no special requirement on the reference cosmology but is should not be too far deviated from the data. Once $\alpha$ and $\beta$ are determined, we can fix them in other cosmology fits. In the Pantheon sample, the corrected apparent magnitude $m_{B,{\textrm{corr}}}^*=m_B^*+\alpha X_1-\beta \mathcal{C}$ are reported. Therefore, we don't need to do the color and stretch corrections any more, so we fix $\alpha=\beta=0$ in equation (\ref{eq:mu_sn}). The statistical uncertainty $\mathbf{D}_{\textrm{stat}}$ and systematic uncertainty $\mathbf{C}_{\textrm{sys}}$ are also given in Ref.\cite{Scolnic:2017caz}. The total uncertainty matrix of distance modulus is given by
\begin{equation}
  \mathbf{\Sigma}_\mu=\mathbf{D}_{\textrm{stat}}+\mathbf{C}_{\textrm{sys}}.
\end{equation}

We convert the distance modulus of SNe to the normalized comoving distance through the relation
\begin{equation}
  d_c=\frac{1}{1+z}\frac{H_0}{c}10^{\frac{\mu-25}{5}}.
\end{equation}
The uncertainty of $d_c$ is propagated from the uncertainties of $\mu$ and $H_0$ using the standard error propagation formula,
\begin{equation}
  \mathbf{\Sigma}_{d_c}=\mathbf{D}_1\mathbf{\Sigma}_\mu \mathbf{D}_1^{\textrm{ T}}+\mathbf{\sigma}_{H_0}^2\mathbf{D}_2\mathbf{D}_2^{\textrm{ T}},
\end{equation}
where $\sigma_{H_0}$ is the uncertainty of Hubble constant, the superscript `\,T' denotes the transpose of a matrix, $\mathbf{D}_1$ and $\mathbf{D}_2$ are the Jacobian matrices,
\begin{eqnarray}
  \mathbf{D}_1&=&{\textrm{diag}}\left(\frac{\ln10}{5}{\bm d}_c\right),\\
  \mathbf{D}_2&=&{\textrm{diag}}\left(\frac{1}{H_0}{\bm d}_c\right),
\end{eqnarray}
where ${\bm d}_c$ is a vector whose components are the normalized comoving distances of all the SNe Ia in Pantheon, and ${\textrm{diag}}({\bm v})$ is the square diagonal matrix with the elements of vector ${\bm v}$ on the main diagonal.

To obtain the comoving distance - redshift relation from the discrete data points, we use the GP method \cite{Seikel:2012uu} to reconstruct the $d_c(z)$ function, and the derivatives $d_c'(z)$ and $d_c''(z)$ can be obtained simultaneously in the reconstruction procedure. The GP can reconstruct a function $y=f(x)$ from the discrete data points $(x_i,y_i$) without assuming a particular parametrisation of the function $f(x)$. It assumes that the data points are drawn from the multivariate Gaussian distribution,
\begin{equation}
  \bm{y}\sim \mathcal{N}(\bm{\mu},\mathbf{K}(\bm{x},\bm{x})+\mathbf{C}),
\end{equation}
where $\bm{x}=\{x_i\}$, $\bm{y}=\{y_i\}$, $\bm{\mu}$ is the mean of the Gaussian distribution, $\mathbf{C}$ is the covariance matrix of the data, $[\mathbf{K}(\bm{x},\bm{x})]_{ij}=k(x_i,x_j)$ is another covariance matrix which controls the behavior of the reconstructed function. All the freedoms of GP originate from the choice of the covariance function $k(x_i,x_j)$. There are several covariance functions available, but any covariance function should be symmetric, positive definite and monotonously decreasing with $|x_i-x_j|$. In this paper, we use the simplest and most widely used squared-exponential covariance function defined by
\begin{equation}
  k(x_i,x_j)=\sigma_f^2\exp\left[-\frac{(x_i-x_j)^2}{2l^2}\right].
\end{equation}
The hyperparameters $\sigma_f$ and $l$ are optimized by maximizing the marginalized likelihood. For more details on the GP, please refer \cite{Seikel:2012uu}.

It has been noticed that the GP method, although can reconstruct the function $f(x)$ itself with a relatively high precision, the reconstructed derivatives, especially the higher order derivatives of $f(x)$ have large uncertainty. If there are observational constraints on the derivatives of the function, then the uncertainty can be reduced. From equation (\ref{eq:E(z)}) we can know that $d'_c(z)=1/E(z)=H_0/H(z)$, thus the direct measurement of Hubble parameters can be used as an additional constraint on the first order derivative of $d_c(z)$. Here we use the 51 $H(z)$ data points (except for $H_0$) complied in Ref.\cite{Magana:2017nfs}, which is, to our knowledge, the largest data sample available at present. These $H(z)$ data points are measured from two different methods, i.e. the differential age of galaxies (DAG) method and the baryon acoustic oscillations (BAO) method. Since the BAO method relies on the cosmological model, to avoid the model dependence we only use the remaining 31 data points measured from the DAG method.

To normalize $H(z)$, the precise measurement of $H_0$ is necessary. It is well known that there is more than $3\sigma$ tension between the values measured from the local distance ladders and that from the global CMB radiation, where the former gives $H_0=73.24\pm 1.74~{  km~s^{-1}~Mpc^{-1}}$ \cite{Riess:2016jrr}, while the later gives $H_0=67.4\pm 0.5~{  km~s^{-1}~Mpc^{-1}}$ \cite{Aghanim:2018eyx}. To investigate the influence of Hubble constant on the reconstruction, we consider these two different values as the prior on $H_0$.

\section{Results and discussions}\label{sec:results}

The publicly available python package \textsf{GaPP} \cite{Seikel:2012uu} is used to do the GP reconstructions. First we do the reconstruction from the Pantheon dataset, then we add the $H(z)$ data to make a combined reconstruction. Since $w(z)$ depends on the matter density parameter $\Omega_M$, we fix it to the value of Planck 2018 results, i.e. $\Omega_M=0.315$ \cite{Aghanim:2018eyx}. The impact of different $\Omega_M$ values on the reconstruction of $w(z)$ will be discussed later. Note that the other quantities ($d_c(z)$, $E(z)$ and $q(z)$) are independent of $\Omega_M$. The absolute magnitude of SNe Ia is degenerated with the Hubble constant, and we fix it to $M_B=-19.35$, the best-fitting value of $\Lambda$CDM model.

\begin{figure}[htbp]
\centering
\includegraphics[width=0.5\textwidth]{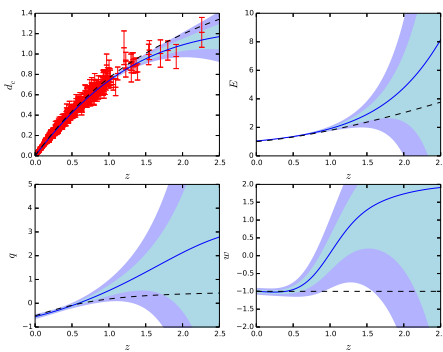}
\caption{\label{fig:H0_67.4_SN} The GP reconstructions of $d_c(z)$, $E(z)$, $q(z)$ and $w(z)$ from SNe, with prior $H_0=67.4\pm 0.5~{km~s^{-1}~Mpc^{-1}}$. The black dashed curves are the predictions of flat $\Lambda$CDM model with parameters $\Omega_M=0.3$, $H_0=68.9~{km~s^{-1}~Mpc^{-1}}$.}
\end{figure}

The GP reconstructions of $d_c(z)$, $E(z)$, $q(z)$ and $w(z)$ from SNe data with prior $H_0=67.4\pm 0.5~{km~s^{-1}~Mpc^{-1}}$ are plotted in Fig.\ref{fig:H0_67.4_SN}. The blue curves are the reconstructed central values, and the shaded regions are the $1\sigma$ and $2\sigma$ uncertainties. For comparison, we also plot the best-fitting curves of $\Lambda$CDM model, with the best-fitting parameters $\Omega_M=0.3$, $H_0=68.9~{km~s^{-1}~Mpc^{-1}}$. Fig.\ref{fig:H0_67.4_SN} shows that the reconstructions of all the four quantities are well matched to the $\Lambda$CDM model in low redshift ($z<0.5$) region. However, in redshift region higher than 0.5, the reconstructed curves show discrepancy from the $\Lambda$CDM model. This discrepancy is especially obvious for $w(z)$, which is more than $2\sigma$ deviation from $-1$ in the intermediate redshift region. Due to the large uncertainties at $z>0.5$, the rest three quantities ($d_c(z)$, $E(z)$ and $q(z)$) are still consistent with the $\Lambda$CDM model within $2\sigma$ confidence level.

\begin{figure}[htbp]
\centering
\includegraphics[width=0.5\textwidth]{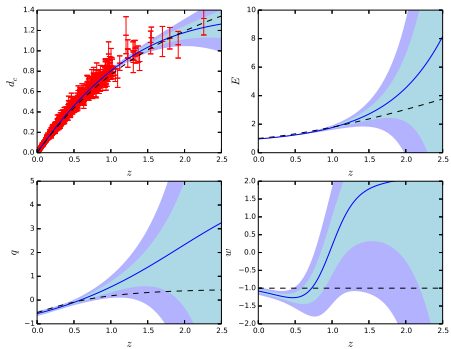}
\caption{\label{fig:H0_73.24_SN} The same to Fig.\ref{fig:H0_67.4_SN} but with prior $H_0=73.24\pm 1.74~{km~s^{-1}~Mpc^{-1}}$.}
\end{figure}

The GP reconstructions from the same data but with prior $H_0=73.24\pm 1.74~{ km~s^{-1}~Mpc^{-1}}$ are plotted in Fig.\ref{fig:H0_73.24_SN}. Similar to Fig.\ref{fig:H0_67.4_SN}, the reconstructions of $d_c(z)$, $E(z)$ and $q(z)$ are consistent with the $\Lambda$CDM model within $2\sigma$ confidence level, especially in the low redshift region, they are excellently in agreement with $\Lambda$CDM. However, the reconstruction of $w(z)$ shows more than $2\sigma$ discrepancy from $-1$ in low redshift region, but it is consistent with $-1$ in the intermediate and high redshift regions. This is contrary to Fig.\ref{fig:H0_67.4_SN}, which shows the discrepancy from $\Lambda$CDM in the intermediate redshift region. This implies that the $H_0$ value has significant impact on the reconstruction of $w(z)$.

\begin{figure}[htbp]
\centering
\includegraphics[width=0.5\textwidth]{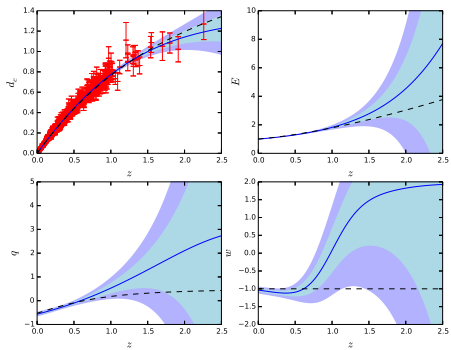}
\caption{\label{fig:H0_none_SN} The same to Fig.\ref{fig:H0_67.4_SN} but with no prior on $H_0$.}
\end{figure}

There is more than $3\sigma$ tension between the two $H_0$ priors we used here. A wrong $H_0$ prior may lead to a wrong result on the reconstruction of $w(z)$. It is interesting to see which of the two $H_0$ priors is more consistent with the SNe Ia data. To this end, instead of reconstructing the normalized comoving distance $d_c(z)$, we first directly reconstruct the un-normalized comoving distance $D_c(z)$ (and its derivatives) as the function of redshift, where
\begin{equation}
  D_c(z)=\frac{c}{H_0}d_c(z)=\frac{1}{1+z}D_L(z).
\end{equation}
Because $D_L(z)$ is directly measured from SNe Ia, $D_c(z)$ is independent of $H_0$. Since $d'_c(0)=1/E(0)\equiv 1$, we can infer the Hubble constant by
\begin{equation}
  H_0=\frac{c}{D'_c(0)}, \sigma_{H_0}=\frac{c\sigma_{D'_c(0)}}{[D'_c(0)]^2},
\end{equation}
where $D'_c(0)$ is the derivative of $D_c(z)$ at $z=0$. Using this method, we obtain $H_0=70.6\pm 0.5~{km~s^{-1}~Mpc^{-1}}$, which is approximately the mean value of the local and global measurements of $H_0$. Then, we normalize $D_c(z)$ (and its derivatives) with the inferred $H_0$ and calculate $E(z)$, $q(z)$ and $w(z)$ in the same way as previous cases. The results are plotted in Fig.\ref{fig:H0_none_SN}. Now the reconstructed $w(z)$ is consistent with $-1$ within $2\sigma$ confidence level in the whole redshift region, except for a small region near $z\sim 1.2$. In addition, from the $q(z)$ subfigure we see a turn point at $z=0.59_{-0.06}^{+0.08}$, where the universe changes from accelerating to decelerating. The location of the turn point is in good agreement with the prediction of $\Lambda$CDM model.

\begin{figure}[htbp]
\centering
\includegraphics[width=0.5\textwidth]{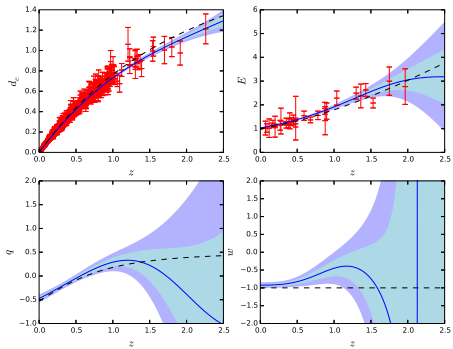}
\caption{\label{fig:H0_67.4_SN_Hz} The GP reconstruction of $d_c(z)$, $E(z)$, $q(z)$ and $w(z)$ from SNe+$H(z)$, with prior $H_0=67.4\pm 0.5~{km~s^{-1}~Mpc^{-1}}$. The black dashed curves are the predictions of flat $\Lambda$CDM model with parameters $\Omega_M=0.3$, $H_0=68.9~{km~s^{-1}~Mpc^{-1}}$.}
\end{figure}
\begin{figure}[htbp]
\centering
\includegraphics[width=0.5\textwidth]{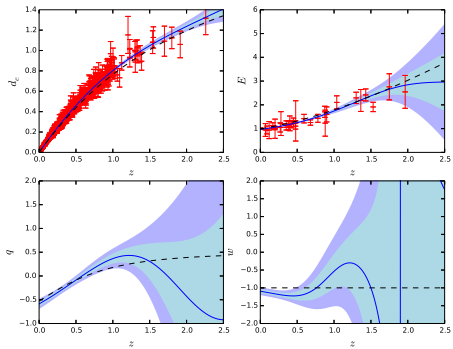}
\caption{\label{fig:H0_73.24_SN_Hz} The same to Fig.\ref{fig:H0_67.4_SN_Hz} but with prior $H_0=73.24\pm 1.74~{km~s^{-1}~Mpc^{-1}}$.}
\end{figure}

The reconstructed quantities, especially $E(z)$, $q(z)$ and $w(z)$ which depend on the derivatives of $d_c(z)$, have large uncertainty in high redshift region. To reduce the uncertainty, we combine the SNe Ia with the 31 DAG $H(z)$ data in the reconstruction, where the inverse of the normalized $H(z)$ data are treated as an additional constraint on the first order derivative of $d_c(z)$. The reconstruction from the SNe+$H(z)$ data with prior $H_0=67.4\pm 0.5~{km~s^{-1}~Mpc^{-1}}$ and prior $H_0=73.24\pm1.74~{km~s^{-1}~Mpc^{-1}}$ are plotted in Fig.\ref{fig:H0_67.4_SN_Hz} and Fig.\ref{fig:H0_73.24_SN_Hz}, respectively. Compared with Fig.\ref{fig:H0_67.4_SN} and Fig.\ref{fig:H0_73.24_SN}, we may see that adding the $H(z)$ data can partially reduce the uncertainty. But now the constructed $d_c(z)$ is not consistent with $\Lambda$CDM within $2\sigma$ confidence level in the intermediate redshift region. The $H_0$ prior directly affects $E(z)$ and thus affects the slope of $d_c(z)$. With the small $H_0$ prior, the slope of $d_c(z)$ is small, so the reconstructed $d_c(z)$ increases slower than the prediction of $\Lambda$CDM. On the contrary, with the large $H_0$ prior, the reconstructed $d_c(z)$ increases faster than the prediction of $\Lambda$CDM. Similar to the SNe only case, with small $H_0$ prior, $w(z)$ is deviated from $-1$ in the intermediate redshift region, while with large $H_0$ prior, $w(z)$ is deviated from $-1$ in the low redshift region.

\begin{figure}[htbp]
\centering
\includegraphics[width=0.5\textwidth]{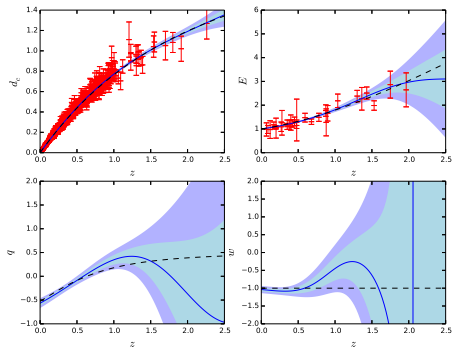}
\caption{\label{fig:H0_none_SN_Hz} The same to Fig.\ref{fig:H0_67.4_SN_Hz} but with no prior on $H_0$.}
\end{figure}

Finally, we reconstruct $d_c(z)$, $E(z)$, $q(z)$ and $w(z)$ without $H_0$ prior using the similar method mentioned above. We directly reconstruct the un-normalized $D_c(z)$, treating the $c/H(z)$ data as an additional constraint on the first derivative of $D_c(z)$. Then we infer $H_0$ and its uncertainty from the reconstructed $D_c(z)$ curve using equation (17). The results are plotted in Fig.\ref{fig:H0_none_SN_Hz}. The inferred Hubble constant is $H_0=70.5\pm 0.5~{km~s^{-1}~Mpc^{-1}}$, which agrees with the value inferred from SNe Ia only. The $d_c(z)$ is excellently in agreement with $\Lambda$CDM model in the whole redshift region, and $E(z)$ is coincident with $\Lambda$CDM model within $1\sigma$ confidence level. Compared with Fig.\ref{fig:H0_none_SN}, an obvious difference can be seen in the $q(z)$ subfigure. Except for an unambiguous accelerating-to-decelerating turn point at $z=0.59_{-0.05}^{+0.05}$, there is another possible, but with large uncertainty turn point near $z\sim 1.8$. In addition, $q(z)$ and $w(z)$ are consistent with $\Lambda$CDM within $2\sigma$ confidence level. Therefore we conclude that the combined data of SNe+$H(z)$ shows no evidence for the deviation from the standard cosmological model.

\begin{figure}[htbp]
\centering
\includegraphics[width=0.5\textwidth]{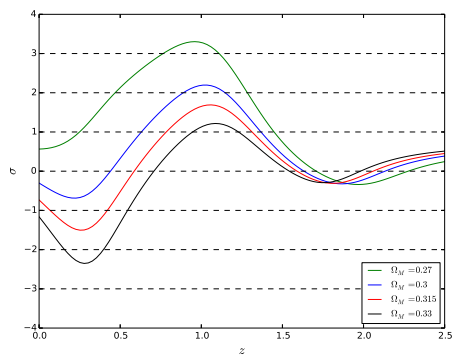}
\caption{\label{fig:sigma} The deviation of $w(z)$ from $-1$ for different $\Omega_M$ values.}
\end{figure}

To investigate the influence of $\Omega_M$ on $w(z)$, we reconstruct $w(z)$ from SNe+$H(z)$ data with different $\Omega_M$ values and without $H_0$ prior. Define the significance of deviation of $w(z)$ from $-1$ as
\begin{equation}
  \sigma(z)=\frac{w-(-1)}{\sigma_{w}},
\end{equation}
where $w$ and $\sigma_w$ are the reconstructed central value and $1\sigma$ uncertainty, respectively. We plot $\sigma(z)$ for different $\Omega_M$ values ($\Omega_M=[0.27,0.30,0.315,0.33]$) in Fig.\ref{fig:sigma}. It is shown that the $\Omega_M$ value has significant effect on the reconstruction of $w(z)$. For $\Omega=0.27$, $w(z)$ deviates from $-1$ at more than $3\sigma$ confidence level in the intermediate redshift region. For $\Omega_M=0.3$ and $\Omega_M=0.33$, $w(z)$ deviates from $-1$ at more than $2\sigma$ confidence level near $z\sim 1.0$ and $z\sim 0.3$, respectively. For $\Omega_M=0.315$, however, $w(z)$ is consistent with $-1$ within $2\sigma$ confidence level in the whole redshift region. It is an interesting feature that the deviation of $w(z)$ from $-1$ reaches its peak value at redshift $z \sim 1.0$ for any $\Omega_m$. Due to the large uncertainty of $w(z)$ in the high redshift region, $w(z)$ is consistent with $-1$ within $1\sigma$ uncertainty for all the $\Omega_M$ values. Therefore, a precise measurement of the matter density parameter $\Omega_M$ is necessary to tightly constrain the dark energy.

It should be noticed that the uncertainty of $E(z)$, $q(z)$ and $w(z)$ increases sharply beyond redshift $z\sim 1.5$ due to the lack of data points. Adding the $H(z)$ data can partially reduce the uncertainty, but it is still unacceptably large. Especially, the $w(z)$ reconstructed from SNe+$H(z)$ has a sudden break and the uncertainty blows up near $z\sim 2$, which is of course unreasonable. This flaw of GP method has already been noticed in Ref.\cite{Seikel:2012uu}. To overcoming this flaw more data points in the high redshift region are needed.

Recently, Zhang and Li \cite{Zhang:2018gjb} used the Union2.1 and JLA compilations of SNe Ia combined with the $H(z)$ data to reconstruct the dark energy in redshift region $z<1.5$. They found that the Union2.1+$H(z)$ and JLA+$H(z)$ data give similar results, i.e., both datasets present a hint of dynamical dark energy, but cannot exclude the constant dark energy. They also investigated the effect of $H_0$ and $\Omega_M$ on the construction, and showed that $H_0$ has notable influence on the results, but the influence of $\Omega_M$ is slight. In our work, with the most recent SNe Ia data, we reconstructed the dark energy up to redshift $z\sim 2.5$ and got similar results to Ref.\cite{Zhang:2018gjb}. However, our results show that both $H_0$ and $\Omega_M$ have an unnegligible effect on the reconstruction of dark energy. This difference may be cause by the reduction of uncertainty at $z<1.5$. From Fig.\ref{fig:sigma}, we see that $\Omega_M$ value only affect the result bellow $z\sim 1.5$. Beyond $z\sim 1.5$, due to the large uncertainty, the influence of $\Omega_M$ is negligible.

\section{Summary}\label{sec:summary}

In this paper, we have reconstructed the EoS of dark energy and the cosmic expansion from SNe Ia using non-parametric method. To improve the significance we have also added the direct measurement of Hubble parameter, $H(z)$ data, to make a combined reconstruction. However, even if with the $H(z)$ data, the reconstruction still has large uncertainty in high redshift region. It is found that the $H_0$ value has strong effect on the reconstruction. Without $H_0$ prior, the inferred Hubble constant from the combination of SNe+$H(z)$ data is $H_0=70.5\pm 0.5~{km~s^{-1}~Mpc^{-1}}$, thus alleviates the tension between the local and global measurements of $H_0$. We have also investigate if the matter density parameter, $\Omega_M$, has some influence on the reconstruction. It is shown that the reconstruction of $w(z)$ strongly depends on $\Omega_M$. With the inferred Hubble constant and the Planck 2018 matter density parameter ($\Omega_M=0.315$), the reconstructed $w(z)$ is consistent with $\Lambda$CDM model within $2\sigma$ confidence level. With current observational accuracy, it is still premature to distinguish between evolving and constant dark energy.

\acknowledgments{This work has been supported by the National Natural Science Fund of China (Grant Nos. 11603005 and 11775038).}

\end{document}